\documentclass[twocolumn,secnumarabic,amssymb,groupaddress,nofootinbib,aps,prl,showpacs,showkeys]{revtex4-1}

\usepackage{amsthm}
\usepackage{amsmath}
\usepackage{amssymb}
\usepackage{color}
\usepackage{adjustbox}
\usepackage{graphicx}
\usepackage{hyperref}
\usepackage{multirow}
\usepackage{array}

\newcommand{\ket}[1]{\left| #1\right. \rangle}

\newcommand{\dd}{\mathrm{d}}

\begin{document}
\title{Quantum nonlocality with arbitrary limited detection efficiency}
\author{Gilles P\"utz}
\email[]{Gilles.Puetz@unige.ch}
\affiliation{Group of Applied Physics, University of Geneva, CH-1211 Geneva 4, Switzerland}
\author{Djeylan Aktas}
\affiliation{Universit\'e Nice Sophia Antipolis, Laboratoire de Physique de la Mati\`ere Condens\'ee, CNRS UMR 7336, Parc Valrose, 06108 Nice Cedex 2, France}
\author{Anthony Martin}
\affiliation{Group of Applied Physics, University of Geneva, CH-1211 Geneva 4, Switzerland}
\author{Bruno Fedrici}
\affiliation{Universit\'e Nice Sophia Antipolis, Laboratoire de Physique de la Mati\`ere Condens\'ee, CNRS UMR 7336, Parc Valrose, 06108 Nice Cedex 2, France}
\author{S\'ebastien Tanzilli}
\affiliation{Universit\'e Nice Sophia Antipolis, Laboratoire de Physique de la Mati\`ere Condens\'ee, CNRS UMR 7336, Parc Valrose, 06108 Nice Cedex 2, France}
\author{Nicolas Gisin}
\affiliation{Group of Applied Physics, University of Geneva, CH-1211 Geneva 4, Switzerland}

\date{\today}



\begin{abstract}
The demonstration and use of nonlocality, as defined by Bell's theorem, rely strongly on dealing with non-detection events due to losses and detectors' inefficiencies. Otherwise, the so-called detection loophole could be exploited. The only way to avoid this is to have detection efficiencies that are above a certain threshold. We introduce the intermediate assumption of limited detection efficiency, that is, in each run of the experiment, the overall detection efficiency is lower bounded by $\eta_{min} > 0$. Hence, in an adversarial scenario, the adversaries have arbitrary large but not full control over the inefficiencies. We analyse the set of possible correlations that fulfill Limited Detection Locality (LDL) and show that they necessarily satisfy some linear Bell-like inequalities. We prove that quantum theory predicts the violation of one of these inequalities for all $\eta_{min} > 0$. Hence, nonlocality can be demonstrated with arbitrarily small limited detection efficiencies. We validate this assumption experimentally via a twin-photon implementation in which two users are provided with one photon each out of a partially entangled pair. We exploit on each side a passive switch followed by two measurement devices with fixed settings. Assuming the switches are not fully controlled by an adversary, nor by hypothetical local variables, we reveal the nonlocality of the established correlations despite a low overall detection efficiency.

\end{abstract}

\maketitle

\textit{Introduction ---} When studying the discoveries in fundamental physics of the past century, one cannot help but come across Bell's seminal work on the nonlocal nature of quantum theory~\cite{Bell1964}. It implies that quantum physics can produce correlations which cannot be explained by a common past with local variables propagating contiguously. This has not only proven fascinating from a foundational point of view, but also given rise to applications in device independent quantum information processing (DIQIP)~\cite{Brunner:RMP}, such as quantum key distribution~\cite{Ekert91,BarrettKent05,Acin06}, randomness generation~\cite{Colbeck2006,Pironio2010}, or entanglement certification~\cite{Bancal11,Barreiro13}. For a semi broad-audience presentation of these concepts, see~\cite{Gisin14}.

\begin{figure}
\includegraphics[width=0.5\textwidth]{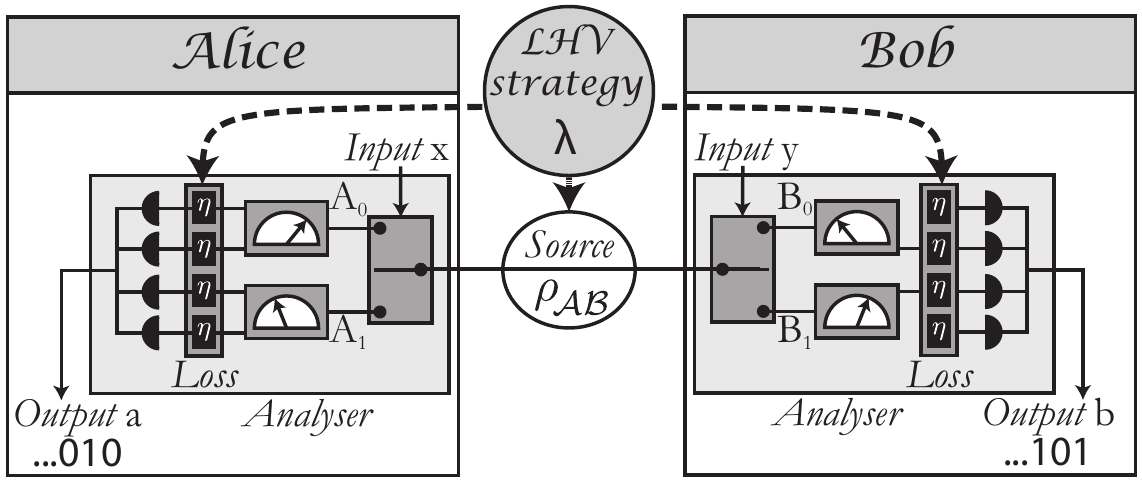}
\caption{Two boxes receive each a particle, emitted by a common source, each. They are given inputs $x$ and $y$, which we depict here as the setting of an active switch, and return outputs $a$ and $b$, respectively. There is the possibility for non-detection events, in which case the corresponding output variable takes the value $\varnothing$. Since these losses can be seen as happening inside the box, they can depend on the inputs $x$ and $y$, respectively. We analyze the limited detection local case of this scenario, meaning that a local hidden variable $\lambda$ not only fully describes the state of the particle, but can also influence whether or not a non-detection event occurs.}
\label{figprinciple}
\end{figure}

Let us briefly recall the concept of local and nonlocal correlations. Assume that a source emits pairs of particles that travel to two distant stations, say Alice and Bob. As depicted in \figurename{~\ref{figprinciple}}, the two experimentalists perform one out of several possible measurements on the individual particles they receive and record the associated outcomes. We denote Alice's and Bob's measurement choices by $x$ and $y$ and their recorded outcomes by $a$ and $b$, respectively. They can then compute the correlation $P(ab|xy)$. Given the setup, it seems natural to think that any correlations that Alice and Bob can observe in this way are due to particles having a common past, as they come from the same source. We refer to this common past by $\lambda$. Correlations that can be explained by the existence of such a parameter are called \textit{local}:
\begin{align}
\label{locality}
P_{L}(ab|xy)=\int\dd\lambda\rho(\lambda)P(a|x\lambda)P(b|y\lambda).
\end{align}
Bell's work shows that there are quantum correlations that cannot be reproduced by such a local model, proving that quantum physics is inherently \textit{nonlocal}. Nonlocality has since been demonstrated many times \cite{Brunner:RMP}, and finds repercussion in a variety of applications~\cite{Ekert91,Acin06,Colbeck2006,Pironio2010,Bancal11}.

However, when demonstrating quantum nonlocality, several issues have to be addressed. Here, we are specifically interested in one of them: what happens if the particles can be lost on the way to or inside the measurement stations, including the possibility that they reach the detectors but are simply not registered. In this case, we say that $a=\varnothing$ or $b=\varnothing$. One immediate idea is of course to carefully analyze why the particles get lost and, if the mechanisms are well understood, to simply discard these cases. This means that Alice and Bob postselect on the cases for which they both register a detection event: $a\neq\varnothing$ and $b\neq\varnothing$. Notably, this opens up the possibility that fully local correlations appear nonlocal if our understanding of the cause of the non-detection events is incorrect, a situation that we wish to avoid~\cite{Branciard11,Pomarico11}. This is especially relevant in the case of active adversaries in DIQIP applications. Another option is to consider the non-detection events as an additional possible outcome and simply check if the resulting correlation is nonlocal. In this case, one will never mistake a local correlation for a nonlocal one. The drawback, however, is that even highly nonlocal distributions may now appear local.

In the end, the only way address this issue, usually called the detection loophole, consists of not only producing highly nonlocal correlations but also having a high enough detection efficiency~\cite{Brunner:RMP}. If the latter is not satisfied, even a perfect state preparation and perfectly calibrated measurement apparatus do not help and one is left with an inconclusive experiment unless the detection loophole is assumed to be not exploited. It is only recently that all the potential loopholes in a Bell test have been closed in a single experiment for the first time~\cite{Hensen15}. However, closing the detection loophole in long-range experiments and applications remains difficult. Namely, after a transmission through $10$ km of the currently best optical fiber, the losses already exceed the tolerable threshold. We therefore take a different route in this paper.

Here, we introduce the concept of limited detection locality. It consists of an intermediate assumption between neglecting the detection loophole and closing it completely. We show that this assumption, even when arbitrarily weak, allows demonstrating nonlocality by postselection with arbitrarily low overall detection efficiency. We then connect the concepts of limited detection locality and measurement dependent locality~\cite{Putz14,Aktas2015}. Finally, we discuss a dedicated twin-photon experiment that we performed, and show how quantum nonlocality can be revealed despite low detection efficiencies without having to make the fair-sampling assumption.


\textit{Limited Detection Locality (LDL)--- } Assume that there exist a fixed $\eta_{min}$ and $\eta_{max}$ with $[\eta_{min},\eta_{max}]\subsetneqq [0,1]$ such that
\begin{align}
\label{lde}
\eta_{min}\leq P(a\neq\varnothing|x\lambda)\leq\eta_{max},
\end{align}
and similarly for Bob. This corresponds to the assumption that, for any input $x$ and any common local variable $\lambda$, there is a probability of at least $\eta_{min}$ and at most $\eta_{max}$ of having a detection. Consider for example a world in which the polarization degree of freedom of photons was as of yet undiscovered. It is nowadays well known that the detection efficiency of most photonic detectors is indeed sensitive to the polarization degree of freedom of light. However the detection efficiency never goes up to $1$ or down to $0$, which corresponds to our assumption of limited detection efficiency with nontrivial $\eta_{min}$ and $\eta_{max}$. It is also worth noting that the assumption can be imposed at the cost of weakening the established correlations by a partial postselection. We do not discuss this further here, but refer to \cite{SuppMat} for an explanation of this idea.
We refer to correlations fulfilling conditions (\ref{locality}) and (\ref{lde}) as \textit{limited detection local}. Note that technically, the case of $[\eta_{min},\eta_{max}]=[0,1]$ can still be analyzed with our model, making it possible to retrieve the results discussed in~\cite{Branciard11}.

In an experiment, one can additionally determine the actual observed detection efficiencies, which may of course be different for the different sets of inputs. It is necessary that some detections occurred for all possible sets of inputs, which reads
\begin{align}
\label{de}
P(a\neq\varnothing|x)=\eta^A_{x}>0.
\end{align}
Since
\begin{align}
P(a\neq\varnothing|x)=\int\dd\lambda\rho(\lambda)P(a\neq\varnothing|x\lambda),
\end{align}
we have that $\eta_{min}\leq\eta^A_x\leq\eta_{max}$. All of this holds analogously for Bob's side. To ease notation, we define $\eta_{xy}=\eta^A_{x}\eta^B_y$.

We can now focus on the postselected limited detection local distributions given by
\begin{align}
P(ab|xy,a\neq\varnothing,b\neq\varnothing)=\frac{P(ab|xy)}{\eta_{xy}}.
\end{align}

Similarly to local correlations, these postselected limited detection local correlations fulfill certain conditions. More precisely, they form a convex polytope and therefore respect a set of linear Bell-like inequalities~\cite{SuppMat}. Making the additional assumption that $\eta_{xy}=\eta_{x'y'}$ for all $x,x',y,y'$, one of these inequalities is, for example, given by
\begin{align}
\label{ldlineq}
\eta_{min}^2P(00|00,a\neq\varnothing,b\neq\varnothing)\nonumber\\
 - \eta_{min}\eta_{max}P(01|01,a\neq\varnothing,b\neq\varnothing)\nonumber\\
-\eta_{min}\eta_{max}P(10|10,a\neq\varnothing,b\neq\varnothing)\nonumber\\
-\eta_{max}^2P(00|11,a\neq\varnothing,b\neq\varnothing)\leq 0.
\end{align}
In an experiment with given losses, the experimentalists can check for which values of $\eta_{min}$ and $\eta_{max}$ their observed correlations violate this inequality. They can then conclude that no limited detection local model with these parameters could have reproduced them.

Interestingly, there are quantum correlations that do not fulfill this inequality independently of the observed detection efficiencies $\eta_{xy}$ and for any upper bound $\eta_{max}$ as long as $\eta_{min}>0$. In fact, this is achieved by all quantum correlations violating Hardy's paradox~\cite{Hardy93} and can therefore be realised using any sufficiently pure partially entangled 2-qubit state with the right set of projective measurements. This may be quite surprising since it is well-known that without the assumption of limited detection efficiency (\ref{lde}), a minimal observed detection efficiency of $\eta=\sum_{xy}\eta_{xy}/4>\frac{2}{3}$ is required to demonstrate nonlocality for 2 parties using binary inputs and outputs. However, making the arbitrarily weak additional assumption that $P(a\neq\varnothing|x\lambda)\geq\eta_{min}>0$ allows demonstrating quantum nonlocality despite arbitrarily large losses and detection inefficiencies.

\textit{Link to measurement dependent locality (MDL)--- } Another way to counterfeit nonlocal correlations using only local resources is if the common history $\lambda$ is correlated with the inputs $x$ and $y$. If the correlation can be arbitrary, then any nonlocal correlation can be counter-fitted in this way, so limitations have to be imposed to be able to make any conclusions. Together with some co-authors, we recently studied the case of measurement dependent local correlations~\cite{Putz14,Aktas2015} that are defined in the following way:
\begin{align}
\label{MDLcorr}
P(abxy)=\int\dd\lambda\rho(\lambda)P(xy|\lambda)P(a|x\lambda)P(b|y\lambda)\\
\label{MDLcond}
\ell\leq P(xy|\lambda)\leq h.
\end{align}
Note that if Alice and Bob each have $N$ inputs, then $0\leq\ell\leq\frac{1}{N^2}\leq h\leq 1$ due to the normalization of probability distributions. Similarly to this paper, we showed that the set of MDL-correlations for fixed $\ell$ and $h$ can be analysed using Bell-like inequalities. 

It turns out that there exists a strong link between the concepts of limited detection locality and measurement dependent locality. We make this connection explicit by the following theorem, whose demonstration is detailed in the Supplemental Material~\cite{SuppMat}:

\textbf{Theorem:} Assume that we have a correlation that can be produced by using a combination of postselected limited detection (\ref{lde}) and measurement dependent (\ref{MDLcond}) local (\ref{MDLcorr}) resources, with parameters $(\eta_{min},\eta_{max})$ and $(\ell,h)$, respectively. Then, this correlation can also be reproduced using only measurement dependent local resources with $\ell'=\frac{\eta_{min}^2}{\eta_{max}^2}\ell$ and $h'=\frac{\eta_{max}^2}{\eta_{min}^2}h$.

Intuitively, the link comes from the fact that an adversary that is allowed, via postselection, to discard unsatisfying inputs is in fact influencing the choice of inputs. A consequence of this theorem is that whenever a correlation cannot be reproduced by a measurement dependent local model with bounds $\ell'$ and $h'$, then it can also not be realized using limited detection efficiencies with $\frac{\eta_{min}^2}{\eta_{max}^2}\geq N^2\ell'$ and $\frac{\eta_{max}^2}{\eta_{min}^2}\leq N^2 h'$, where $N$ is the number of inputs for each of the two parties. This allows us to use any result derived for the MDL-scenario and apply them to LDL-correlations. Even more interestingly, we are now able to deal with the problems of losses and measurement dependence in a straightforward way since we can simply focus exclusively on measurement dependence.

\begin{figure}
\includegraphics[width=0.5\textwidth]{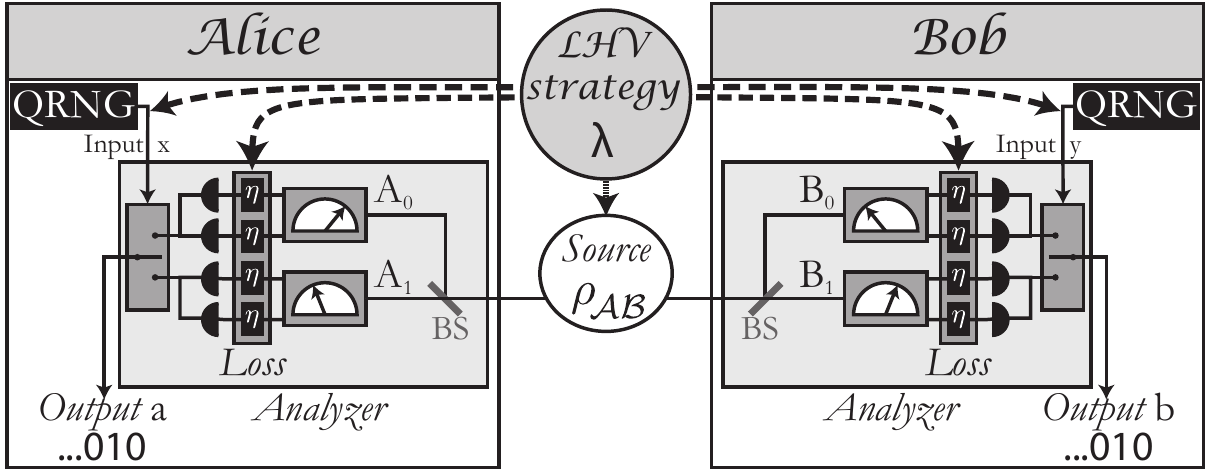}
\caption{A source emits supposedly entangled photons towards two measurement stations. In each station, a passive switch routes the individual incoming photons to one out of two pairs of detectors, corresponding to either settings $0$ or $1$. The boxes are given inputs $x$ and $y$, which decide which pairs of detectors are switched on, and return outputs $a$ and $b$ depending on which detector fires. If the photon goes towards the inactive pair of detectors it is simply considered as lost and not registered. We compare this scenario to the scenario where a local hidden variable $\lambda$ that can influence the losses and settings try to emulate our observed correlations.}
\label{figexp}
\end{figure}

\textit{Experiment:} We now discuss the twin-photon based experimental realization whose principle is depicted in \figurename{~\ref{figexp}}. Each user's station comprises 2 measurement devices with fixed settings, \textit{i.e.}, 4 single-photon detectors (8 in total). The detection events at Alice's are space-like separated from the corresponding ones at Bob's. In order to emulate active switches between different analyzis settings at each user's location without being limited by driving speeds, we exploit instead a simple and elegant solution for passive implementation, referred to as Zbinden switches~\cite{Gisin99}. First, standard 50/50 beam-splitters connect the incoming photons to the measurement devices. Then, two quantum random number generators (QRNG) provide the inputs $x$ for Alice and $y$ for Bob to randomly select one pair of detectors associated with one measurement. The other two detectors of each box remain inactive. Consequently, when, in any run of the experiment, a photon hits an inactive detector, it is merely disregarded. Hence, with perfect devices, the maximum attainable detection efficiency is $50\%$.

Our source produces partially entangled photon pairs close to the state
\begin{align}
\label{goldenstate}
\left|\Psi\right. \rangle = \frac{1}{\sqrt{3}}\left(\frac{\sqrt{5}-1}{2}\left| 00\right. \rangle+\frac{\sqrt{5}+1}{2}\left|11 \right. \rangle\right),
\end{align}
and the 8 detectors are set up to perform measurements close to the ideal projective measurements $\ket{A_0(\theta)}=\cos\theta\ket{0}+\sin\theta\ket{1}$, $\ket{A_1(\theta)}=\mid A_0(\theta-\frac{\pi}{4})\rangle$, $\ket{B_0(\theta)}=\ket{A_0(-\theta)}$ and $\ket{B_1(\theta)}=\ket{A_1(-\theta)}$, with $\theta=\arccos{\sqrt{\frac{1}{2}+\frac{1}{\sqrt{5}}}}$. Further experimental details are given in~\cite{Aktas2015} as well as in the Supplemental Material~\cite{SuppMat}. Without any additional noise, the resulting correlations would violate inequality (\ref{ldlineq}), $\forall\eta_{min}>0$ and $\forall\eta_{max}$. Obviously, due to noise and imperfections even in the postselected case this is not fully realized. Our postselected correlations, meaning that we discard all the non-detection events, can be found in Table~\ref{probexp}. While we did not precisely determine the overall detection efficiency in our experiment, it is worth to note that it stands clearly below the required $\frac{2}{3}$ efficiency that is necessary to close the detection loophole in the case of 2 parties with binary inputs and outputs~\cite{Brunner:RMP,Hensen15}. However, assuming that the losses are the same for all input pairs, \textit{i.e.}, $\eta_{xy}=\eta_{x'y'}$, we can still conclude that we reveal quantum nonlocality for $\frac{\eta_{min}}{\eta_{max}}>0.267$. This follows from inequality (\ref{ldlineq}).

There are two ways to illustrate this result. First, assume that a local hidden variable guides, on each side, the particle through the beam-splitter and influences the detection efficiency. Such a local hidden variable would need to have the power to reduce the overall detection efficiency down to below $26.7\%$ in order to mimic nonlocal correlations. Next, consider an adversary partially controlling the detectors' efficiencies. Assuming standard 50/50 beam-splitters, we have that $\eta_{max}\leq 0.5$ and one gets $\eta_{min}=0.134$, \textit{i.e.}, such an adversary would have to lower the detector efficiencies corresponding to settings they do not like down to below $13.8\%$ in order to have local correlations appear nonlocal. In our experiment, due to the beam-splitters, the losses in the fibres and the difficulties of perfectly aligning 8 detectors, we can safely consider that $\eta_{max}\leq 0.1$, leading to a required $\eta_{min}\leq 0.027$ in order to reproduce the correlations using a local hidden variable model.

\begin{table}
\begin{tabular}{c c c l l|l l}
& & & \multicolumn{4}{c}{Bob}\\
& & &\multicolumn{2}{c|}{y=0}&\multicolumn{2}{c}{1}\\
&  \rotatebox[origin=c]{90}{x}&  \rotatebox[origin=c]{90}{a} &  \multicolumn{1}{c}{b = 0}  & \multicolumn{1}{c|}{1}& \multicolumn{1}{c}{0} & \multicolumn{1}{c}{1}\\
\multirow{4}{*}{\rotatebox[origin=c]{90}{Alice}}&\multirow{2}{*}{\rotatebox[origin=c]{90}{0}}& \rotatebox[origin=c]{90}{0} & $0.01977(12)$&$0.01003(9)$&$0.00435(6)$&$0.00112(3)$\\
& & \rotatebox[origin=c]{90}{1}& $0.00830(8)$&$0.19763(36)$&$0.01576(3)$&$0.24282(41)$\\
\cline{2-7}
&\multirow{2}{*}{\rotatebox[origin=c]{90}{1}} & \rotatebox[origin=c]{90}{0} &$0.02536(15)$&$0.00325(5)$&$0.00089(3)$&$0.03583(19)$\\
& & \rotatebox[origin=c]{90}{1} &$0.00106(3)$&$0.23631(41)$&$0.01411(12)$&$0.18349(39)$
\end{tabular}
\caption{Measured unconditional probabilities $P(abxy)$, including the distribution of the inputs.}
\label{probexp}
\end{table}

Additionally, the way our experiment was built and carried out clearly shows the link between limited detection efficiency and measurement dependence. At both Alice's and Bob's stations, the respective beam-splitter sends the photon to either the detectors corresponding to input $0$ or to the ones corresponding to input $1$. The inputs $x$ and $y$ determine which pair of detectors we consider as switched on each side. If the photon goes towards the detectors for input $1$ while we input $0$, then the detection is not registered and the photon is lost. The non-detection events thereby allow the photons to 'choose' the input they will reply to. The limited detection assumption then corresponds to limiting this choice, \textit{i.e.}, while the action of the beam-splitter can be biased, it cannot be fully biased. The link to measurement dependence is here obvious.

We can use theorem 1 to extend our analysis to the case where we consider our random number generators providing the inputs $x$ and $y$ to be biased. Let us assume that there is some measurement dependence as defined in (\ref{MDLcond}), given by parameters $(\ell,h)$, as well as some limited detection efficiency, given by parameters $(\eta_{min},\eta_{max})$. Using an MDL inequality provided in \cite{Putz14} and theorem 1, we then conclude that our experiment shows quantum nonlocality for
\begin{align}
\frac{\ell}{h}\left(\frac{\eta_{min}}{\eta_{max}}\right)^4>0.15529.
\end{align}

\textit{Conclusion --- } Losses and detection inefficiencies have been serious obstacles in all experiments dedicated to test Bell's theorem. This is why a loop-hole free Bell-test took so long to be successfully demonstrated. They are also part of the main weak points that an adversary may attack in any task whose security relies on quantum nonlocality. To help dealing with both of these issues from a theoretical point of view, we have introduced the additional assumption of limited detection efficiency (\ref{lde}). The assumption is that the inefficiencies in the setup are only partially exploited, an idea that we consider very intuitive in itself especially in the case of a fundamental Bell test, provided Nature is non-malicious.
The case of an adversary is also interesting. If one assumes an adversary with full control over Alice's and/or Bob's devices, then one needs to resort to full Device-Independent Quantum Information Processing (DIQIP)~\cite{Brunner:RMP,Pironio09}. However, semi-DIQIP is extremely appealing and timely as it allows one to lower the extremely high requirement of full DIQIP. For example, considering DIQIP as a way to check the proper functioning of our system, \textit{i.e.}, going back to the original idea of self-testing~\cite{Mayers04}, it is very natural to assume that possible defects do not set $\eta_{min}=0$. In such a case, our approach provides much simpler self-tests. Even against an active adversary, there are cases in which it is reasonable to assume that they have only partial control on the detection efficiencies.

Making the fair-sampling assumption when it is not warranted can lead to disastrously wrong conclusions~\cite{Pomarico11,Gerhardt_fake_PRL_2011,Romero_Fair_NJP_2013}. The same holds for the LDL assumption in cases where it is not justified. However, due to the fact the LDL assumption is strictly weaker than assuming fair-sampling, it can clearly hold and be justified more often. In other words, if an adversary is not able to fully suppress the detection events, \textit{i.e.}, $\eta_{min}>0$, nonlocality can be demonstrated and exploited.

We connected the ideas of limited detection locality and measurement dependent locality. We showed that results from studying measurement dependent local correlations can be applied to the case of limited detection locality. Moreover, it is possible to deal with detection efficiencies and lack of measurement independence at the same time, which we hope will be of use for future Bell experiments.

Finally, we carried out an experiment to bring these theoretical ideas to the real world. Our twin-photon based demonstration, having detection efficiencies too low to close the detection loophole~\cite{Hensen15}, remains conclusive as long as we assume a limited detection efficiency such that $\frac{\eta_{min}}{\eta_{max}}>0.267$. It furthermore reveals very clearly the link between measurement dependence and limited detection. We believe that the methods depicted here will be of use to any frontier experiments or applications that do not manage to close the detection loophole completely.

\textit{Note added --- } During the review process, we became aware, in addition to Ref.~\cite{Hensen15}, of two more experimental demonstrations related to loophole-free Bell tests~\cite{Zeilinger2015,Sae2015}.

\textit{Acknowledgments --- } We acknowledge Alex May who wrote down an early version of a proof similar to the one given in Appendix 2 showing a connection between limited detection and measurement dependence, as well as Roger Colbeck, who mentioned the idea of limited detection in private discussions. We also thank Valerio Scarani for discussions and Laurent Labont\'e for his help in the data acquisition process. We further acknowledge financial support from the Universit\'e Nice Sophia Antipolis, the European project CHIST-ERA DIQIP, the Swiss NCCR-QSIT, the Fondation iXCore pour la Recherche, as well as the Fondation Simone \& Cino Del Duca (Institut de France).


\newpage
\onecolumngrid
\appendix

\section{Polytopal structure of Limited Detection Local correlations}
Consider the case where $N$ parties perform a nonlocality experiment. The input and outcome of the $i$-th party will be denoted by $X_i$ and $A_i$ respectively. We consider the case where non-detection events can occur, they will be denoted by $A_i$ taking the value $\varnothing$. We will denote by $A'_i$ the outcome of party $i$ after postselecting on having a detection. As discussed in the main text, we make the assumption of limited detection locality and show that these correlations form a polytope. The theorem stated here is more general than needed for the main text, where we only consider the case of 2 parties.\\

\textit{Definitions:} Let $\{A_i\}_{i=1}^N$, $\{A'_i\}_{i=1}^N$, $\{X_i\}_{i=1}^N$ be sets of random variables with alphabets $\{1\cdots m_i,\varnothing\}$, $\{1\cdots m_i\}$ and $\{1\cdots n_i\}$ respectively, $m_i,n_i\in\mathcal{N}$. In the following, the corresponding lower case letters will denote values in the respective alphabet. We will denote probability distributions over a random variable $V$ by $P_V$, the value of this distribution for a given value of $V$ by $P_V(v)$. For ease of notation, we will often omit the random variable and just write $P(v)$. We will denote conditional probability distributions over a random variable $V$ conditioned on a random variable $W$ by $P_{V|W}$. In the case of continuous random variable we denote the probability density by $\rho_V$. In the following we assume that all the probability distributions are well defined. The set of all probability distributions over $V$ will be denoted by $\mathcal{P}_V$ and of all conditional probability distributions over $V$ conditioned on $W$ by $\mathcal{P}_{V|W}$.\\

We define the following sets:

\begin{itemize}
\item The sets of 1-party distributions with limited detection:
\begin{align*}
\mathcal{LD}_i(\eta_{min},\eta_{max})=\Big\{ P_{A_i|X_i}\in\mathcal{P}_{A_i|X_i} : 1-\eta_{min}\leq P_{A_i|X_i}(\varnothing|x)\leq 1-\eta_{max} \forall x\in\{0\cdots n_i\} \Big\}
\end{align*}
\item The set of $N$-party limited detection local distributions:
\begin{align*}
\mathcal{LDL}_N(\{\eta_{min,i}\}_{i=1}^N,\{\eta_{max,i}\}_{i=1}^N) =\Big\{P_{A_1\ldots A_N|X_1\ldots X_N}\in&\mathcal{P}_{A_1\ldots A_N|X_1\ldots X_N} : \exists \Lambda \text{ s.t. } \\
&P(a_1\ldots a_N|x_1\ldots x_N) = \int\dd\lambda\rho(\lambda)\prod_{i=1}^N P(a_i|x_i\lambda), \\
&P_{A_i|X_i \Lambda=\lambda}\in \mathcal{LD}_i(\eta_{min,i},\eta_{max,i}) \forall\lambda, \forall i\Big\}
\end{align*}
\item The set of $N$-party postselected limited detection local distributions:
\begin{align*}
\mathcal{LDLPS}_N(\{\eta_{min,i}\}_{i=1}^N,\{\eta_{max,i}\}_{i=1}^N,\{\eta_{x_1\ldots x_N}\}_{x_1\ldots x_N})=\Big\{&P_{A'_1\ldots A'_N|X_1\ldots X_N}\in\mathcal{P}_{A'_1\ldots A'_N|X_1\ldots X_N} : \\
&\exists Q_{A_1\ldots A_N|X_1\ldots X_N}\in\mathcal{LDL}(\{\eta_{min,i}\}_{i=1}^N,\{\eta_{max,i}\}_{i=1}^N), \\
&Q(a_1\neq\varnothing\ldots a_N\neq\varnothing|x_1\ldots x_N)=\eta_{x_1\ldots x_N},\\
&P(a'_1\ldots a'_N|x_1\ldots x_N) = \frac{Q(a'_1\ldots a'_N|x_1\ldots x_N)}{\eta_{x_1\ldots x_N}}\Big\}.
\end{align*}
\end{itemize}

We further define the following two sets, which we will prove to be the vertices of $\mathcal{LD}_i(\eta_{min},\eta_{max})$ and $\mathcal{LDL}_N(\{\eta_{min,i}\}_{i=1}^N,\{\eta_{max,i}\}_{i=1}^N)$:
\begin{align*}
\mathcal{V}^{\mathcal{LD}_{i}}(\eta_{min},\eta_{max})=\Big\{ V_{A_i|X_i}\in&\mathcal{P}_{A_i|X_i} : \forall x\in\{1\ldots n_i\}\exists! a_x\in\{1\ldots m_i\} \text{ and }\eta_x\in\{\eta_{min},\eta_{max}\} \text{ s.t.}\\
&V(a_x|x)=\eta_x \text{, } V(\varnothing|x)=1-\eta_x \text{ and otherwise } V(a|x)=0 \Big\}
\end{align*}
\begin{align*}
\mathcal{V}^{\mathcal{LDL}_N}(\{\eta_{min,i}\}_{i=1}^N,\{\eta_{max,i}\}_{i=1}^N) =\Big\{V_{A_1\ldots A_N|X_1\ldots X_N}\in&\mathcal{P}_{A_1\ldots A_N|X_1\ldots X_N} : \\
&\exists V_i\in\mathcal{V}^{\mathcal{LD}_{i}}(\eta_{min,i},\eta_{max,i}) \text{ s.t. }\\
&V(a_1\ldots a_N|x_1\ldots x_N) = \prod_{i=1}^NV_i(a_i|x_i)\Big\}
\end{align*}

With these definitions, we can now state the theorem. It refers to polytopes, which for the purposes of this work are simply seen as a convex structure with a finite set of vertices. Equivalently they can be defined by a finite set of inequalities.\\

\textbf{Theorem:} For fixed $N$, $\{\eta_{min,i}\}_{i=1}^N$ and $\{\eta_{max,i}\}_{i=1}^N$, $\mathcal{LDL}_N(\{\eta_{min,i}\}_{i=1}^N,\{\eta_{max,i}\}_{i=1}^N)$ is a polytope whose vertices are a subset of $\mathcal{V}^{\mathcal{LDL}_N}(\{\eta_{min,i}\}_{i=1}^N,\{\eta_{max,i}\}_{i=1}^N)$. Furthermore, for fixed $\{\eta_{x_1\ldots x_N}\}_{x_1\ldots x_N})$, $\mathcal{LDLPS}_N(\{\eta_{min,i}\}_{i=1}^N,\{\eta_{max,i}\}_{i=1}^N,\{\eta_{x_1\ldots x_N}\}_{x_1\ldots x_N})$ is also a polytope.\\

\textit{Proof:} To ease notation we will omit writing $N$, $\{\eta_{min,i}\}_{i=1}^N$ and $\{\eta_{max,i}\}_{i=1}^N$ from now on. The first part of the theorem follows from the following two lemmas.\\

\textbf{Lemma 1:} $\mathcal{LD}_i$ is a polytope with vertices given by $\mathcal{V}^{\mathcal{LD}_{i}}$.\\

\textit{Proof:} Due to the normalisation of probability distributions, \textit{i.e.},
\begin{align*}
\sum_{a_i}P(a_i|x_i)=1 \text{ }\forall x_i,
\end{align*}
we have that $P(\varnothing|x_i)=1-\sum_{a_i=1}^{m_i}P(a_i|x_i)$ and we can therefore work in the lowerdimensional subspace given by $a_i\in\{1\ldots m_i\}$. In this subspace, we are then left with the polytope defined by the inequalities $\eta_{min}\leq\sum_{a_i=1}^{m_i}P(a_i|x_i)\leq\eta_{max}$. This is the definition of a hypercube whose vertices are defined by the corresponding part of $\mathcal{V}^{\mathcal{LD}_i}$.\\

The Lemma follows. $\square$\\

\textbf{Lemma 2:} Let $\mathcal{Q}$ and $\mathcal{R}$ be polytopes with vertices $\mathcal{V_Q}$ and $\mathcal{V_R}$ respectively. Let $\mathcal{S}=\Big\{S: \exists\Lambda\text{ s.t. } S(u,v)=\int\dd\lambda\rho(\lambda)Q_\lambda(u)R_\lambda(v) \text{ with }Q\in\mathcal{Q},R\in\mathcal{R}\Big\}$. Let $\mathcal{V_S}=\Big\{V_S : V_S(u,v)=V_Q(u)V_R(v) \text{ with }V_Q\in\mathcal{V_Q}, V_R\in\mathcal{V_R}\Big\}$.

Then $\mathcal{S}$ is a polytope whose vertices are a subset of $\mathcal{V_S}$.\\

\textit{Proof:} By definition, $\mathcal{S}$ is convex and $\mathcal{V_S}\in\mathcal{S}$.

Let $S\in\mathcal{S}$, then by definition we have:
\begin{align*}
S(u,v)&=\int\dd\lambda\rho(\lambda)Q_\lambda(u)R_\lambda(v)\\
&=\int\dd\lambda\rho(\lambda)\sum_{i}q_{\lambda,i}V_Q^{i}(u)\sum_{j}r_{\lambda,j}V_R^{j}(v)\\
&=\sum_{(ij)}\Big(\int\dd\lambda\rho(\lambda)q_{\lambda,i}r_{\lambda,j}\Big)V_Q^{i}(u)V_R^{j}(v)\\
&=\sum_{(ij)}s_{ij}V_S^{ij}(u,v)
\end{align*}
In the first step we use the fact that any element of $\mathcal{Q}$ and $\mathcal{R}$ can be written as a convex combination of their vertices and we define $q_{\lambda,i}\geq 0$, $\sum_i q_{\lambda,i}=1$ and $r_{\lambda,j}\geq 0$, $\sum_j r_{\lambda,j}=1$. In the last step we defined $s_{ij}=\int\dd\lambda\rho(\lambda)q_{\lambda,i}r_{\lambda,j}$, which fulfils $s_{ij}\geq 0$ and $\sum_{ij}s_{ij}=1$ and also used the definition of $V_S^{ij}$.\\

This proves the Lemma.$\square$\\

Using these two Lemmas in conjunction (and using Lemma 2 iteratively) proves that $\mathcal{LDL}$ is a polytope and that $\mathcal{V_LDL}$ contains its vertices.

To finalize the proof of the theorem we need to show that $\mathcal{LDLPS}$ is a polytope as well. This can be seen directly since the set is obtained by slicing $\mathcal{LDL}$ with the hyperplanes defined by $P(a_1\neq\varnothing\ldots a_N\neq\varnothing|x_1\ldots x_N)=\eta_{x_1\ldots x_N}$. Cutting a polytope with hyperplanes results in another polytope. The final step is a simple rescaling of the entries (equivalent to rescaling the axes) and therefore the set remains again a polytope. \\

This proves the theorem. $\square$\\

\section{A more general method of dealing with detection inefficiencies--- } 
It is possible to impose any desired $\eta_{min}$ at the price of adding some noise to the system. Assume that Alice and Bob set their detection systems, which we assume to have an efficiency $\eta$, such that any time a non-detection event occurs, the system still gives an outcome with the probability $\eta_{min}$. In this way, Alice and Bob impose that their detection systems have a limited detection efficiency given by the chosen $\eta_{min}$ and $\eta_{max}=1$ and they can treat the resulting correlations by the tools presented above. This however comes at the price of adding local noise to their correlations. In fact, assume that Alice and Bob would share the nonlocal correlation $P_{NL}$ if the detectors were perfect and there were no losses (\textit{e.g.}, given by projective measurements on a pure quantum state) and denote by $P_{NL}^A$ and $P_{NL}^B$ the marginal distributions of Alice and Bob, respectively. In the non-detection cases, the detection systems are set up such that they give with probability $\eta_{min}$ an outcome given by the local distributions $P_L^A$ and $P_L^B$, respectively. Then, by postselecting on the cases where the detection systems gave an outcome, Alice and Bob share the correlation:
\begin{align}
P=\Big(\eta^2P_{NL}+\eta(1-\eta)\eta_{min}(P_{NL}^AP_L^B+P_{L}^AP_{NL}^B)\nonumber\\
+(1-\eta)^2\eta_{min}^2P_L^AP_L^B\Big)\frac{1}{(\eta+(1-\eta)\eta_{min})^2}.
\end{align}
They can then analyze this correlation using the tools of limited detection efficiency presented above.

One possibility of dealing with losses and detector inefficiencies is to assign the non-detection events to an additional outcome and treat the resulting correlations using the usual tools of nonlocality. This corresponds exactly to the strategy we just presented with $\eta_{min}=1$. However, our approach is more general, allowing to assign only a fraction of the non-detection events to an outcome and postselecting on the rest. For a fixed detection efficiency $\eta$, our method therefore encompasses both of the previous strategies, full postselection and full assignment to an additional outcome, and additionally allows for an arbitrary mixture of the two. It is at this point not obvious to us that for a given experiment (meaning for a given $P_{NL}$ and a given $\eta$), all of these strategies would yield the same result. We leave it up for future works to analyze this question in more detail.

\section{Limited Detection Locality and Measurement Dependent Locality}
In this section we prove the link between limited detection local and measurement dependent local distributions. This can be proven more generally, here we only present the 2-party version.\\

\textit{Definitions:} We introduce the random variables $D_A$ and $D_B$ with alphabet $\{0,1\}$ such that $D_A=0$ if and only if $A=\varnothing$. We define the set of limited detection local distributions allowing for measurement dependence:
\begin{align*}
\mathcal{MDLDL}(\ell,h,\eta_{min},\eta_{max})=\big\{P_{AD_ABD_BXY} : &P(ad_Abd_Bxy)=\int\dd\lambda\rho(\lambda)P(xy|\lambda)P(ad_Abd_B|xy\lambda), \\
 & \eta_{min}\leq P_{D_AD_B|XY\Lambda}(11|xy\lambda)\leq\eta_{max},\\ 
 &P(ad_Abd_B|xy\lambda)=P(ad_A|x\lambda)P(bd_B|y\lambda)\\
&\ell\leq P(xy|\lambda)\leq h,\\
 &\int\dd\lambda\rho(\lambda)=1, \rho(\lambda)\geq 0\big\}
\end{align*}

We also define the set of measurement dependent local correlations:
\begin{align*}
\mathcal{MDL}(h,\ell)=\big\{P_{ABXY} : &P(abxy)=\int\dd\lambda\rho(\lambda)P(xy|\lambda)P(ab|xy\lambda), \\
 &P(ab|xy\lambda)=P(a|x\lambda)P(b|y\lambda)\\
&\ell\leq P(xy|\lambda)\leq h,\\
 &\int\dd\lambda\rho(\lambda)=1, \rho(\lambda)\geq 0\big\}
\end{align*}

\textit{Theorem:} If 
\begin{align*}
P_{AD_ABD_BXY}&\in \mathcal{MDLDL}(\ell,h,\eta_{min},\eta_{max})
\end{align*}
then
\begin{align*}
P_{ABXY|D_A=1,D_B=1}\in \mathcal{MDL}(\frac{\eta_{min}}{\eta_{max}}\ell,\frac{\eta_{max}}{\eta_{min}}h).
\end{align*}

\textit{Proof:}
We have
\begin{enumerate}
\item $P(ad_Abd_Bxy)=\int\dd\lambda\rho(\lambda)P(xy|\lambda)P(ad_Abd_B|xy\lambda)$
\item $P(ad_Abd_B|xy\lambda)=P(ad_A|x\lambda)P(bd_B|y\lambda)$
\item $\eta_{min}\leq P_{D_AD_B|XY\Lambda}(11|xy\lambda)\leq\eta_{max}$
\item $\ell\leq P(xy|\lambda)\leq h$.
\end{enumerate}

Let us prove a few implications:
\begin{itemize}
\item If $(AD_A|X)$ and $(BD_B|Y)$ are local, meaning that they fulfil condition 2, then $(D_A|X)$ and $(D_B|Y)$ are also local:
\begin{align*}
P(d_Ad_B|xy\lambda)&=\sum_{a,b}P(ad_Abd_B|xy\lambda)\\
&=\sum_{ab}P(ad_A|x\lambda)P(bd_B|y\lambda)\\
&=P(d_A|x\lambda)P(d_B|y\lambda).
\end{align*}
\item If $(AD_A|X)$ and $(BD_B|Y)$ are local, then $(A|D_AX)$ and $(B|D_BY)$ are also local:
\begin{align*}
P(ab|xd_Ayd_B\lambda)&=\frac{P(ad_Abd_B|xy\lambda)}{P(d_Ad_B|xy\lambda)}\\
&=\frac{P(ad_A|x\lambda)}{P(d_A|x\lambda)}\frac{P(bd_B|y\lambda)}{P(d_B|y\lambda)}\\
&=P(a|xd_A\lambda)P(b|yd_B\lambda).
\end{align*}
\item Knowing less cannot result in knowing more, meaning that upper and lower bounds on $P(\mu|\nu\sigma)$ also hold for $P(\mu|\nu)$: Assume $P(\mu|\nu\sigma)\leq h$, then
\begin{align*}
P(\mu|\nu)&=\sum_{\sigma}P(\sigma)P(\mu|\nu\sigma)\\
&\leq h\sum_\sigma P(\sigma)\\
&=h
\end{align*}
where we used that $\sum_\sigma P(\sigma)=1$. The same holds for lower bounds $\ell\leq P(\mu|\nu\sigma)$. Due to this, condition 3 implies 
\begin{align*}
\eta_{min}\leq P_{D_AD_B|\Lambda}(11|\lambda)\leq\eta_{max}.
\end{align*}
\end{itemize}
Using the implications above, we can show that the conditions imply bounds on $P(xy|D_A=1,D_B=1,\lambda)$:
\begin{align*}
P(xy|D_A=1,D_B=1,\lambda)=\frac{\overbrace{P(D_A=1,D_B=1|xy\lambda)}^{\eta_{min}\leq\cdots\leq\eta_{max}}}{\underbrace{P(D_A=1,D_B=1|\lambda)}_{\eta_{min}\leq\cdots\leq\eta_{max}}}\underbrace{P(xy|\lambda)}_{\ell\leq\cdots\leq h}\\
\Rightarrow \frac{\eta_{min}}{\eta_{max}}\ell\leq P(xy|D_A=1,D_B=1,\lambda)\leq\frac{\eta_{max}}{\eta_{min}}h.
\end{align*}

We can now prove the theorem:
\begin{align*}
P(abxy|D_A=1,D_B=1)&=\int\dd\lambda\rho(\lambda|D_A=1,D_B=1)P(xy|\lambda,D_A=1,D_B=1)P(ab|xy\lambda,D_A=1,D_B=1)\\
&= \int\dd\lambda\rho(\lambda|D_A=1,D_B=1)\overbrace{P(xy|\lambda,D_A=1,D_B=1)}^{\frac{\eta_{min}}{\eta_{max}}\ell\leq\cdots \leq\frac{\eta_{max}}{\eta_{min}}}\cdot\\
&\quad \quad\quad P(a|x\lambda,D_A=1)P(b|y\lambda,D_B=1)
\end{align*}
This is by definition an MDL-correlation:
\begin{align*}
P(ABXY|D_A=1,D_B=1)\in\mathcal{MDL}(\frac{\eta_{min}}{\eta_{max}}\ell,\frac{\eta_{max}}{\eta_{min}}h)
\end{align*}

\section{Experimental details}

\begin{figure}
\includegraphics[width = 0.9\columnwidth]{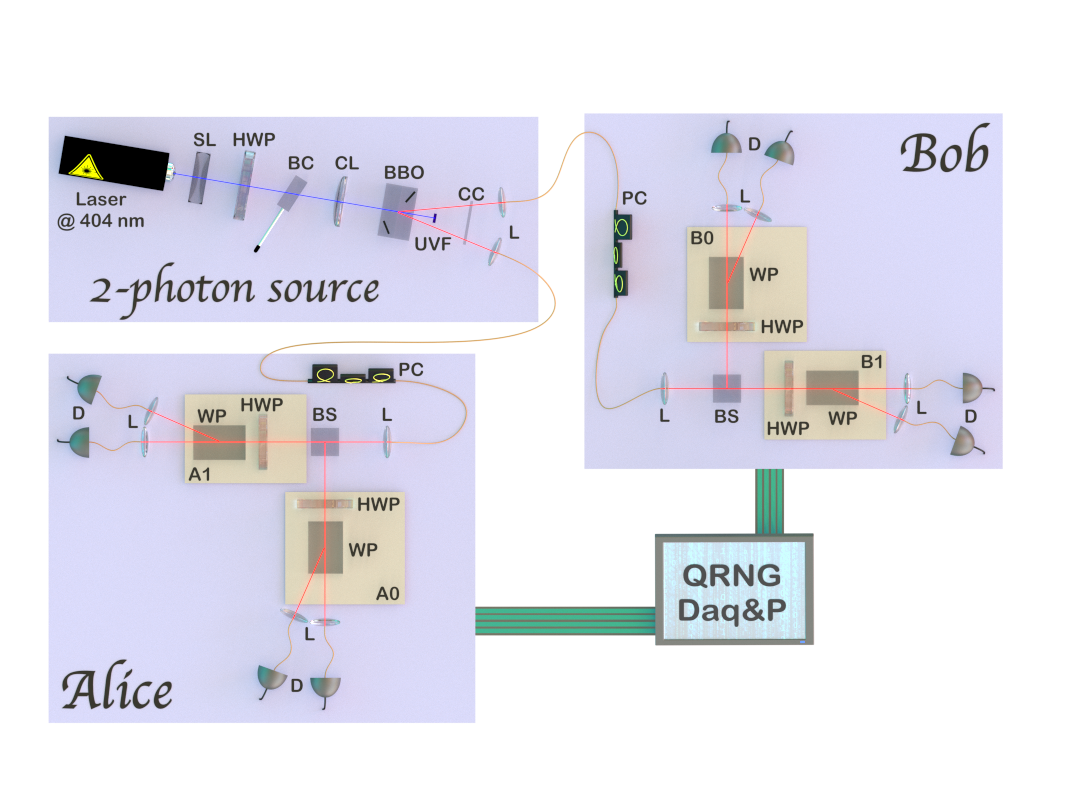}
\caption{\label{fig_source}
Experimental setup - SL: spherical lenses; HWP: half-wave plate; BS: birefringent crystal; CL: cylindrical lens; BBO: beta-barium-borate crystals; UVF: ultra-violet filter; CC: compensation crystal; L: lens; PC: polarization controller; BS: 50/50 beam-splitter; WP: Wollaston prism; D: single photon detector; QRNG: Quantum random number generator; Daq\,\&\,P: Data acquisition \& processing. } 
\end{figure}

To produce the twin-photon partially polarisation entangled state, we employ the source depicted in the top-left hand side of \figurename{\,\ref{fig_source}}. A pump laser at 404\,nm is sent to two cascaded type-I BBO crystals to produce photon pairs at 808\,nm in a coherent superposition of two orthogonal polarisations via the process of spontaneous parametric down-conversion. To generate the twin-photon state of eq.~(10) in the maintext, we set the desired weights of the coherent superposition and the phase between $\ket{VV}$ and $\ket{HH}$ by placing a half wave-plate and a tilted birefringent crystal in the path of the pump laser. The generation process introduces some distinguishability between the two components of the state which is erased by an adapted compensation crystal. Moreover, to remove the discernibility in terms of spatial modes (k-vectors), the paired photons are coupled, each, into a single mode fibre. Thanks to this configuration, we can produce the desired state with a fidelity of 99\% (see also Ref.~\cite{Aktas2015} for more details). 

To perform the measurements, Alice and Bob have each two polarization analysers with fixed settings, defining two bases, \{A$_{\rm 0}$,\,A$_{\rm 1}$\} and \{B$_{\rm 0}$,\,B$_{\rm 1}$\}, respectively. These analysers are made of a half wave-plate and a Wollaston prism. Each prism of the setup has an extinction ratio of 6\,OD. They are connected to the source via a 50/50 beam-splitter.
To choose the measurement basis at both locations, two independent random bit sequences are utilized. This way, only the two randomly selected outputs associated with the given basis are considered, the other two being disregarded~\cite{Gisin99}. This configuration therefore requires the use of 8 single photon detectors, being all Silicon avalanche photodiodes. Alice employs 4 Excelitas SPCM-AQR-16-FC featuring $\sim$60\% detection efficiency and dark count rates of $\sim$100\,Hz, while Bob has 4 QuTools featuring $\sim$50\% detection efficiency and dark count levels of $\sim$2000 Hz. Note that the detectors' configuration is chosen in such a way that it can satisfy, as much as possible, the assumption $\eta_{x,y} = \eta_{x',y'}$ for all $x$, $y$, $x'$, and $y'$ with an overall efficiency of about 5\% (fibre coupling and propagation losses included). The detection arrival times of all the detection events are recorded by a time-to-digital converter (not shown) and the random choice of the basis is applied during the data post-processing thanks to two different quantum random number generators.


%
\end{document}